\documentclass[prl,aps,amssymb,showpacs,twocolumn]{revtex4}
\usepackage{amsmath}
\usepackage{amssymb}
\usepackage{amsthm}
\usepackage{amsfonts}
\usepackage{enumerate}
\usepackage{latexsym}
\usepackage{bm}
\usepackage{graphicx}
\usepackage{subfigure}

\newcommand{\beq}{\begin{equation}}
\newcommand{\eneq}{\end{equation}}

\input{epsf}

\begin{document}

\tolerance 10000

\newcommand{\vk}{{\bf k}}


\title{Topological Entanglement and Clustering of Jain Hierarchy States}

\author{N. Regnault$^1$,  B.A. Bernevig$^{2,3}$, F.D.M. Haldane$^2$ }

\affiliation{$^1$ Laboratoire Pierre Aigrain, Departement de Physique, ENS, CNRS, 24 rue Lhomond, 75005 Paris,  France}
\affiliation{$^2$ Department of Physics, Princeton University, Princeton, NJ 08544}
\affiliation{$^3$ Princeton Center for Theoretical Science, Princeton, NJ 08544}

\begin{abstract}
We obtain the clustering properties and part of the structure of zeroes of the Jain states at filling $\frac{k}{2k+1}$: they are a direct product of a Vandermonde determinant (which has to exist for any fermionic state) and a bosonic polynomial at filling $\frac{k}{k+1}$ which vanishes when $k+1$ particles cluster together.  We show that all Jain states satisfy a ``squeezing rule'' (they are ``squeezed polynomials'') which severely reduces the dimension of the Hilbert space necessary to generate them. The squeezing rule also proves the clustering conditions that these states satisfy. We compute the topological entanglement spectrum of the Jain $\nu=\frac{2}{5}$ state and compare it to both the Coulomb ground-state and the non-unitary Gaffnian state. All three states have very similar ``low energy'' structure. However, the Jain state entanglement ``edge'' state counting matches both the Coulomb counting as well as two decoupled $U(1)$ free bosons, whereas the Gaffnian edge counting misses some of the ``edge'' states of the Coulomb spectrum. The spectral decomposition as well as the edge structure is evidence that the Jain state is universally equivalent to the ground state of the Coulomb Hamiltonian at $\nu=\frac{2}{5}$. The evidence is much stronger than usual overlap studies which cannot meaningfully differentiate between the Jain and Gaffnian states. We compute the entanglement gap and present evidence that it remains constant in the thermodynamic limit. We also analyze the dependence of the entanglement gap and overlap as we drive the composite fermion system through a phase transition.
\end{abstract}

\date{\today}


\maketitle

The experimentally observed fractional quantum Hall (FQH) states in the lowest Landau level (LLL) are thought to be described by Laughlin \cite{laughlin1983} and  hierarchy states modeled by Jain's composite fermion wavefunctions \cite{jain1989}. Jain's original model states have dramatically large overlap with the true Coulomb ground states but the process of flux attachment and projection to the LLL renders them hard to analyze (Monte-Carlo methods have been devised\cite{kamilla1997} for treating variants of the Jain states where
the projection to the LLL is modified, or simply omitted).

The decomposition of Jain's model states into Slater determinants has not been obtained for $N>10$ particles \cite{dev1992}, and (unlike the Laughlin states) they have not been characterized as unique ground states of some model Hamiltonian.  Moreover, their observed large overlap with the ground-states of LLL systems with realistic Coulomb interactions is only empirically understood; this has become most evident recently, when other states,
with identical filling (and ``shift'') \cite{simon2006,bernevig2007},
as the Jain states, but exhibiting different topological order, have been found to have competitive overlaps with the true Coulomb ground-states \cite{regnault2008}. Although these new states are conjectured to represent gapless critical points \cite{read2008}, their large overlap with the Jain states (thought to be gapped in their interior) underscores the need to better understand FQH states from a theoretical standpoint.
In this Letter we describe a previously-unrecognized ``clustering property'' of the Jain model states which allows them to be (partially) characterized as zero-modes of certain pseudo-potential Hamiltonians.
However,  (unlike the Laughlin states), they are \textit{not} unique maximum-density zero modes; while the zero-mode property is insufficient to
completely determine the structure of Jain's model wavefunctions, it provides
a powerful constraint that enables their numerical construction at significantly larger $N$.

The key technical advance reported here is the identification of the structure of Jain states as ``squeezed polynomials'', which means they contain only many-body free-particle configurations obtainable from a ``root'' configuration by a two body operation called ``squeezing'', defined below. This drastically reduces the Hilbert space dimension and allows the generation of Jain states with roughly twice the number of particles previously obtained.

Armed with this technique, we then investigate the topological entanglement spectrum \cite{li2008} of the first state in the hierarchy, the $\nu= 2/5$ Jain state for up to $N=16$ particles, and compare it with both the Coulomb ground-state and the so-called ``Gaffnian'' state, related to a non-unitary conformal field theory (CFT)\cite{simon2007}, which has a Jack polynomial description\cite{bernevig2007}. We find a virtually identical ``low energy'' structure in the Schmidt spectral decomposition of these three states, consistent with the large overlap\cite{simon2006,bernevig2007} of the Jain and Gaffnian states with the Coulomb ground-state. Although the Gaffnian state is very close in both overlap and spectral decomposition to Coulomb and Jain, we directly identify the ``edge'' mode structure of the Coulomb entanglement spectrum and show that it matches the Jain state edge structure as well as that of two $U(1)$ free bosons. We compute the entanglement gap, and show evidence that it remains finite in the the thermodynamic limit. The entanglement gap can be destroyed by tuning the $\nu=2/5$ FQH Coulomb state through a phase transition. We can however, make no definitive statement on the issue of whether the non-unitary Gaffnian state is gapped or gapless \cite{read2008}.

Any fermionic state in the LLL can be written as a product of a Vandermonde determinant and a symmetric polynomial;  for conceptual simplicity,
we will focus on the bosonic variants of model FQH states which omit the
Vandermonde factor (it is straightforward to later go back to the fermionic
states, as multiplication by the Vandermonde factor converts symmetric ``squeezed'' polynomials to antisymmetric ones).
We represent an angular momentum partition $\lambda$ with length $\ell_{\lambda} \le
N$ as a (bosonic) occupation-number configuration $n(\lambda)$ =
$\{n_m(\lambda),m=0,1,2,\ldots\}$ of each of the LLL orbitals $\phi_m(z) = (2\pi m! 2^m)^{-1/2} z^m \exp(-|z|^2/4)$ with angular momentum $L_z = m \hbar$, where, for $m > 0$, $n_m(\lambda)$ is the
multiplicity of $m$ in $\lambda$. It is useful to identify the
``dominance rule'' \cite{stanley1989} (a partial ordering of
partitions $\lambda > \mu$) with the ``squeezing
rule''\cite{sutherland1971} that connects configurations
$n(\lambda)$ $\rightarrow$ $n(\mu)$: ``squeezing'' is a two-particle
operation that moves a particle from orbital $m_1$ to $m_1'$ and
another from $m_2$ to $m_2'$, where $m_1 < m_1' \le m_2' < m_2$, and
$m_1+m_2$ = $m_1'+m_2'$; $\lambda > \mu$ if $n(\mu)$ can be derived
from $n(\lambda)$ by a sequence of ``squeezings''. An interacting LLL polynomial $P_{\lambda}$ indexed by a \emph{root partition}
$\lambda$ is defined as a ``squeezed polynomial'' if it can be expanded in occupation-number non-interacting states (monomials) of orbital occupations $n(\mu)$ obtained
by squeezing on the root occupation $n(\lambda)$:
\begin{equation}
P_{\lambda} = m_\lambda + \sum_{\mu<\lambda} v_{\lambda \mu} m_\mu .
\label{dominantpolynomial}
\end{equation}
\noindent The $v_{\lambda \mu}$ are \emph{rational} number
coefficients. Partitions
$\lambda$ can be classified by $\lambda_1$, their largest part. When
any $P_{\lambda}$ is expanded in monomials $m_\mu$, no orbital with $m
>\lambda_1$ is occupied. $P_\lambda$ can be interpreted as states on a sphere surrounding a monopole with
charge $N_{\Phi}=\lambda_1$\cite{haldane1983}. A large number of FQH states are squeezed polynomials\cite{bernevig2007}. The groundstate wavefunctions of the
Read Rezayi (RR) $Z_k$ sequence\cite{read1999} are Jack polynomials (Jacks) of root occupation
$n(\lambda^0(k,2))=[k0k0k...k0k]$ and Jack parameter
$\alpha_{k,r}=-{(k+1)}/{(r-1)}$ \cite{bernevig2007}. All the Jacks are known to be squeezed polynomials \cite{stanley1989}.
 For the Jacks, the coefficients
$v_{\lambda \mu}$ are explicitly known by recursion
\cite{stanley1989}.

The root configuration of a squeezed polynomial $n(\lambda)$ has the largest variance: $\Delta \lambda = \sum_{i,j=1}^N (\lambda_i -\lambda_j)^2$ of all the partitions $\mu \le \lambda$. A generic state, for example the ground-state of the  Coulomb Hamiltonian in the LLL at some arbitrary filling, has non-zero weight on \emph{all} many-body non-interacting states squeezed from $n(\lambda_{\text{Generic State}} )= \left[\frac{N}{2} 0 0 ...0 0 \frac{N}{2}\right]$ (which has the maximum possible variance), and hence in this case the squeezing property is neither meaningful nor useful. However, for most ``model'' FQH states, the existence of a root configuration drastically reduces the Hilbert space necessary for generating the state and implies many other special properties of the state.

\begin{figure}
\includegraphics[width=3.5in, height=2.1in]{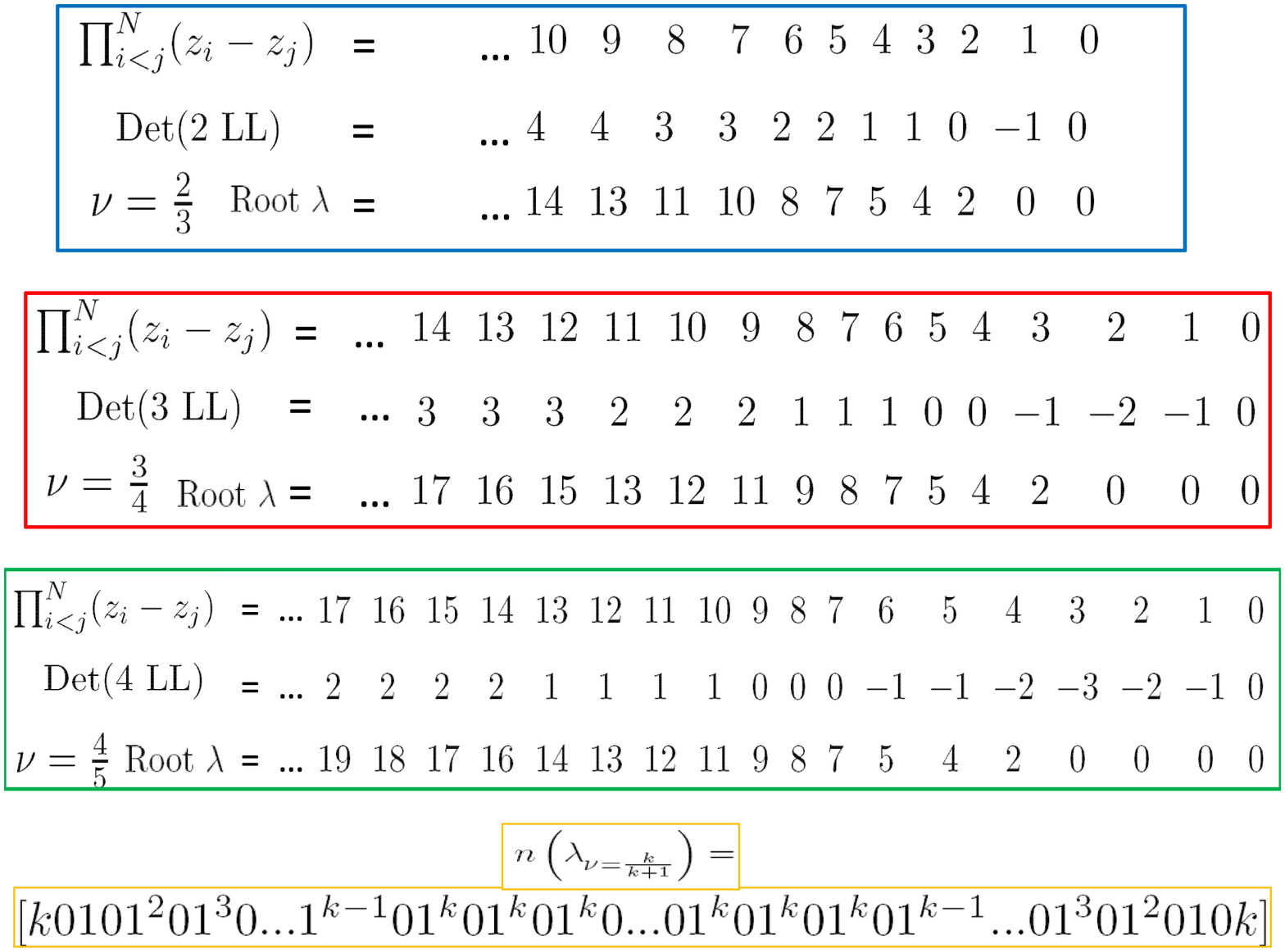}
\caption{Root partition in angular momentum basis for $\nu=\frac{2}{3},\frac{3}{4},\frac{4}{5}...\frac{k}{k+1}$ states can be written as the sum of the Vandermonde determinant partition (single Slater determinant) plus the maximum root partition of the Determinant operator of $k$ Landau levels projected to the LLL. The root occupation configuration contains $k$ particles in $k+1$ orbitals $[1^k0]$ when deep in the bulk. Close to the North and South pole there are deviations from this rule, as shown. }\label{rootconfig}
\end{figure}

We now find the root configuration for all bosonic Jain states at filling $k/(k+1)$, defined as the usual composite fermion states at filling $k/(2k+1)$ divided by a Vandermonde determinant.  We start with the simplest of these states, the $\nu=2/3$ state, defined by placing $N/2$ quasiparticles in the Laughlin $1/2$ state. The operator that implements this is Jain's operator on the plane for $t$ number of quasiparticles \cite{jain1989}:
\begin{equation}
\psi_{\text{t qp}}^J =\text{Det} \left(%
\begin{array}{ccc}
  \partial_1 & \cdots & \partial_N \\
  z_1 \partial_1 & \cdots & z_N \partial_N \\
  \vdots & \cdots & \vdots \\
  z_1^{t-1} \partial_1 & \cdots & z_N^{t-1} \partial_N \\
  1 & \cdots & 1 \\
  z_1 & \cdots & z_N \\
  \vdots & \cdots & \vdots \\
  z_1^{N-t-1} & \cdots & z_N^{N-t-1} \\
\end{array}%
\right) \prod_{i<j}^N (z_i-z_j) \label{jainstate}
\end{equation}
\noindent For any $t$, the above state is not an $\vec{L}=0$ state, and as such, it cannot be the ground-state at $\nu=2/3$ ($t=N/2$), contrary to claims in the literature. The proper composite fermion construction involves writing down the above operator on the sphere and then constructing the state by stereographic projection. However, Eq.(\ref{jainstate}) is sufficient to allow the determination of the root configuration, which, per our definition, is  the maximum variance configuration of orbital occupation number in Eq.(\ref{jainstate}). The Vandermonde factor $\prod_{i<j}^N (z_i-z_j)$ is a single Slater determinant of fermionic root configuration $n(\lambda_0)= [111...111]$ or $\lambda_0 = (N-1, N-2,N-3,N-4,...,6,5,4,3,2,1,0)$; one immediately recognizes in $\lambda_0$ the powers (angular momentum) of the $z_i$ in the Slater determinant. The determinant operator in Eq.(\ref{jainstate}), however, has derivative terms, which we denote by $\partial/\partial z = -1$; its root partition in angular momentum basis is $\lambda_{\text{Det}} = (N-t-1, N-t-1,..., 4,4,3,3,2,2,1,1,0,0,-1)$. There are two states at each angular momentum in $\lambda_{\text{Det}}$ because both $z^m$ and $z^{m+1} \partial/\partial z$ operators contained in the determinant have the same angular momentum $m$. Since the determinant operator now acts on the Vandermonde determinant $\lambda_0$, we could immediately add the two angular momentum partitions, but doing this blindly would cause a problem: the resulting partition $\lambda$, as it describes a polynomial wavefunction $\psi_{\text{t qp}}$ must have all its components positive (the final polynomial must be analytic in $z$'s). As such, the last component of $\lambda_{\text{Det}}$ \emph{cannot} add to the last component of $\lambda_0$; adding these two together would correspond to taking the partial derivative $-1  \rightarrow  \partial/\partial z$ of a constant $0 \rightarrow z^0$, and the result would vanish. As such, the next maximum variance angular momentum partition one can build is $\lambda = (N-1, N-2,N-3,N-4,...4,3,2,1,0)+ (N-t-1, N-t-1,..., 4,4,3,3,2,2,1,1,0,-1,0) = (...14,13,11,10,8,7,5,4,2,0,0)$, where we have written only the angular momentum close to the north pole in the final partition. When written in occupation number, the root configuration is: $n(\lambda) = [201011011011011011...]$. Creating an $\vec{L}=0$ state requires that the north pole be identical to the south pole, and hence the root configuration number for the $\nu=2/3$ state reads: $n(\lambda_{\nu=\frac{2}{3}}) = [201011011011...0110110110102]$. The bulk occupation configuration contains $2$ particles in $3$ orbitals ($110$), as expected for a $\nu= 2/3$ state. For the fermionic state at $\nu= 2/5$: $\psi_{\nu=2/5} =\psi_{\nu=\frac{2}{3}} \cdot \prod_{i<j}^N (z_i-z_j)$, the root occupation number reads $n(\lambda_{\nu=2/5})=[11001001010010100101...10100101001010010011]$.

\begin{figure}
\includegraphics[width=3.5in]{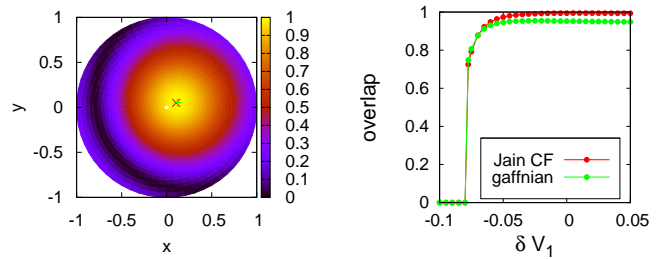}
\caption{{\it Left panel} : Top view of the hemisphere where each point $(x,y,z)$ is associated to one state of the squeezed subspace at $N=12, N_\phi=26$ with components $(x,y,z)$ (squeezed subspace dimension being 3). The color code is the overlap between the Coulomb ground state and the squeezed state associated to the point on the hemisphere. North pole is the Gaffnian, the red cross is the Jain state and the green cross is the point which maximizes the overlap with the Coulomb ground state. {\it Right panel} : Overlaps of both Gaffnian and Jain states with the Coulomb groundstate as a function of added hard-core interaction $\delta V_1$ for $N=16$ particles. A phase transition occurs where the overlap collapses close to $\delta V_1 \simeq -0.08$}\label{overlaps}
\end{figure}

We can obtain the root occupation number for all Jain states at filling $k/(k+1)$. We explicitly show the case $k=3$, with the generalization being trivial.  At $\nu=3/4$, the Jain state is created by attaching one flux to each electron in $3$ occupied LL and then projecting to the LLL:
\begin{equation}
\psi_{\nu=\frac{3}{4}}^J = \text{Det} \left(
  \begin{array}{ccc}
    \partial_1^2 & ... &\partial_N^2\\
    z_1 \partial_1^2 & ...& z_N \partial_N^2\\
    : & :& :\\
    \partial_1 & ... &\partial_N\\
    z_1\partial_1 & ...&z_N \partial_N \\
    : & :&: \\
    1 & ... &1 \\
    z_1 & ...&z_n\\
    : & ...& :\\
  \end{array}
\right) \prod_{i<j} (z_i-z_j)
\end{equation}
 To obtain the maximum variance partition, we again write the Vandermonde determinant $\lambda = (...7,6,5,4,3,2,1,0)$ where we have written only the angular momentum of orbitals close to the north pole. The determinant operator now has $3$ operators of identical angular momentum, because $z^m, z^{m+1}\partial z, z^{m+2} \partial_z^2$ all have identical angular momentum $m$. Moreover, the determinant operator also contains $2$ operators which lower the angular momentum by $1$ unit: $\partial_z$ and $z \partial_z^2$, as well as $1$ operator which lowers the angular momentum by $2$ units: $\partial_z^2$. The maximum variance non-vanishing partition is then $\lambda_{\text{Det}}= (...5 , 5 , 5 , 4 , 4, 4 , 3 , 3 , 3 , 2 , 2 , 2 , 1 , 1 , 1 , 0 , 0 , -1 , -2 , -1 , 0)$. Adding the Vandermonde occupation numbers gives an orbital occupation root configuration of $n(\lambda_{\nu=3/4}) =[3010110111011101110111....011101110110103]$. This procedure and the root configuration for the $\nu=4/5$ state, as well as the general $k$ result are given in Fig.[\ref{rootconfig}].

The root configuration presented in Fig.[\ref{rootconfig}] for general $k$ allows us to determine at least part of the Hamiltonian for which the Jain states are exact zero modes. Cluster $k$ particles at one point, which, by translational invariance (which all FQH ground-states must satisfy), we pick to be the origin. Because \emph{all} the monomials included in the Jain state are squeezed from $n\left(\lambda_{\nu=k/(k+1)}\right)$, placing $k$ particles at the origin results in monomials squeezed from $[00101^201^30...1^{k-1}01^k01^k...1^k01^k01^{k-1}...01^301^2010k]$. These monomials are proportional to $\prod_{i=k+1}^N z_i^2$ and therefore the full polynomial wavefunction vanishes when a $k+1$'th particle is brought at the origin; since the origin is not special by translational invariance, we have:
\begin{equation}
\psi^J_{\nu=\frac{k}{k+1}} (z_1=Z,...,z_k=Z,z_{k+1},...,z_N) \sim \prod_{i=k+1}^N (Z- z_i)^2 \nonumber
\end{equation}
\noindent $\psi^J_{\nu=k/(k+1)}$ are zero modes of the pseudopotential $V_{k+1}^0$ which eliminates the zero angular momentum state of a $k+1$ body cluster; they can be built out of Read-Rezayi $Z_k$ states upon the addition of quasiholes.

\begin{figure}
\includegraphics[width=3.5in]{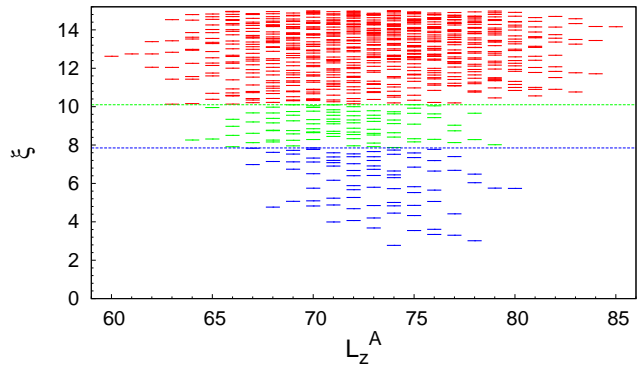}
\caption{Topological entanglement of the pure Coulomb groundstate (no hard-core potential added) at filling $\nu=\frac{2}{5}$. Although the entanglement gap is smaller than the ones reported for the Laughlin and Moore-Read states, one can distinguish the topological entanglement of both the Gaffnian (Jack - blue) and the Jain state in the ``low energy'' (green) structure of the Coulomb ground state. The levels below the light blue line are almost identical to the Gaffnian levels whereas the levels below the green line are almost identical to those of the Jain state, to within $0.003 -3\%$.}\label{entcoulomb}
\end{figure}

Unfortunately, the above Hamiltonian and root partition do not uniquely define the Jain states. Uniform (ground) states on the sphere satisfy the conditions $L^+\psi$ = 0 (highest
weight, HW) and $L^-\psi$ = 0 (lowest weight, LW) where $L^+$ =
$E_0$, and $L^-$ = $N_{\Phi}Z-E_2$, where $Z$ $\equiv$ $\sum_i z_i$,
and $E_n$ = $\sum_iz_i^n\partial/\partial z_i$. Imposing the highest weight condition on the squeezed polynomial with the Jain root partition $n\left(\lambda_{\nu=k/(k+1)}\right)$ results in several linearly independent $\vec{L}=0$ polynomials. We pick the simplest state at bosonic $\nu = 2/3$ or fermionic $\nu=2/5$ to analyze further. From now on, we return to the \emph{fermionic} state. We conjecture that the dimension of the $\vec{L}=0$ subspace of squeezed polynomials with root occupation $n(\lambda_{\nu= 2/5})$ with $N$ electrons is $E[(N+2)/4]$ where $E[x]$ is the integer part of $x$. We have hence reduced the problem of determining the Slater decomposition of a Jain state to the problem of determining $E[(N+2)/4]$ constants rather than the order $N!$ number of constants of each separate Slater determinant. While the usual Monte-Carlo (MC) integration procedures would fail to accurately compute the full decomposition, they may be used to determine the components of the Jain state on this reduced basis. With this method, we are able to obtain the Jain state for up to $N=16$ particles on the sphere geometry (dimension of the squeezed Hilbert space is 99608768, compared to the original full size 155484150); the previous largest size was $N=10$ particles \cite{dev1992}.  Since each component on the $E[(N+2)/4]$ states has its own MC error, several tests have been performed to test the accuracy of this procedure. The overlap between the Jain state we generate using this technique and the corresponding analytical Jain state is higher than 0.9999. One can slightly modify each component within its MC error bar and see how this affects various computed quantities. Thus for $N=14$, the typical relative error on the Coulomb energy is lower than $10^{-5}$ while the one on the entanglement gap is lower than $10^{-2}$. Notice that within the $\vec{L}=0$ subspace of squeezed polynomials, the Jain state is not the best approximation to the Coulomb ground state, but is very close to it as depicted in Fig.[\ref{overlaps}]. We also tried to use this method to obtain higher-order Jain states. Unfortunately, the dimension of the polynomial space squeezed from $n\left(\lambda_{\nu=k/(k+1)}\right)$, while still much smaller with respect to the total number of Slater determinant coefficients, is nonetheless too large for an accurate decomposition of the states.

\begin{figure}
\includegraphics[width=3.5in]{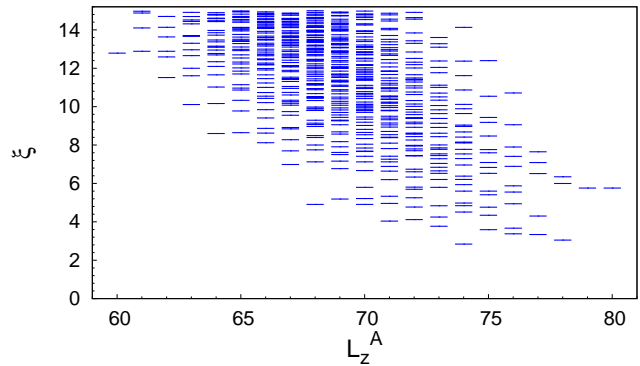}
\caption{Topological entanglement spectrum of the Gaffnian state. The counting of the levels satisfies the generalized Pauli principle of not more than 2 particles in 5 consecutive orbitals and not more than 1 particle in 1 orbital.}\label{entgaff}
\end{figure}

We also constructed the non-unitary Gaffnian state \cite{simon2007} for $N=16$ particles (the squeezed Hilbert space dimension is 91736995), uniquely defined as the highest weight squeezed polynomial with bosonic root occupation $n(\lambda_{\text{Jack}\;\; \nu=2/3}) = [2002002002...2002002]$  \cite{bernevig2007} multiplied by a Vandermonde determinant. In Fig[\ref{entgaff}] we present the overlap of both the Gaffnian and the Jain state with the ground-state of Coulomb plus delta function $\delta V_1$ interaction obtained by exact diagonalization. The overlap is above $95\%$ for both states for $\delta V_1 >-0.06$. There is a phase transition at around $\delta V_1 = -0.08$ and the overlap with both the Jain and the Gaffnian wavefunctions drops dramatically.

\begin{figure}
\includegraphics[width=3.5in]{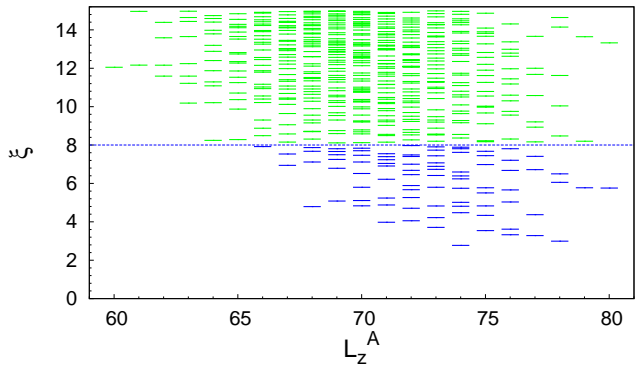}
\caption{Topological entanglement spectrum of the Jain $\nu= 2/5$  state for $N=16$ particles. The ``low energy'' structure of the Jain state (in blue) is almost identical both qualitatively and quantitatively to that of the Gaffnian (below blue line). Because the Jain state is not a pure CFT, it has an entanglement gap of its own with respect to the Gaffnian state.}\label{entjain}
\end{figure}

In order to better understand the remarkably large overlap, as well as to identify the topological order in the Jain and Gaffnian ground state, we compute the topological entanglement spectrum of these states. We place our states on the sphere (comparison is possible because they have identical filling and shift), and cut the state into two hemisphere blocks $A$ and $B$. Following \cite{li2008}, we introduce the "entanglement
spectrum" $\xi$ as $\lambda_i = exp(-\xi_i)$, where $\lambda_i$ are the eigenvalues
of the reduced density matrix $\rho_A$ of one hemisphere. The
eigenvalues can be classified by the number of fermions
$N_A$ in the A block, and also by the total "angular momentum"
$L_z^{(A)}$ of the A block. It was argued \cite{li2008} that
the low-lying spectrum $\xi_i$ of the reduced density matrix
for fixed $N_A$, plotted as a function of $L^{(A)}_z$, should
display a structure reflecting the CFT describing the edge physics. In Fig.[\ref{entcoulomb}], this CFT
spectrum
is defined as every $\xi$ below the light blue line. For interactions at
which the FQH state provides a good description of the
physics, the CFT spectrum should be separated by a gap
from a higher ``non-CFT'' part of the spectrum. This was shown to be the case for the $\nu=5/2$ state \cite{li2008} as well as for the Laughlin $\nu=1/3$ state \cite{zozulya2008}. In our case, $\nu=2/5$, the entanglement gap is not extremely apparent. The three states, Gaffnian, Jain and Coulomb, have the same ``low energy'' entanglement structure as can be seen in Fig.[\ref{entcoulomb}], Fig.[\ref{entgaff}] and Fig.[\ref{entjain}]. The counting of entanglement eigenvalues for the Gaffnian at a certain angular momentum $L_z^{(A)}$ is easily seen to correspond to the counting of occupation number configurations of angular momentum $L_z^{(A)}$ satisfying generalized Pauli principle of \cite{bernevig2007} applied to fermionic states: not more than $2$ particles in $5$ consecutive orbitals and, by virtue of being fermions, not more than $1$ particle in each orbital. The counting of ``edge modes'' reads $1,1,3,5,10...$, same as that obtained by different methods in \cite{bernevig2007A}. The Jain state has a very similar ``low energy'' entanglement structure with the Gaffnian state, but also exhibits extra higher energy levels not present in the Gaffnian. We should remark that the Jain state, not being a pure CFT state (\textit{i.e.}, not obtained as a correlator of CFT primary fields, but rather of their derivatives \cite{hansson2007}), has an entanglement gap of its own. This means that some of the spectral levels present in the Jain entanglement spectrum are non-generic and should become clearly gapped in the thermodynamic limit. For example the entanglement spectrum at the maximum $L_z^{(A)}=80$ is formed by one low-lying eigenvalue and other high-energy ones with very little weight in the Jain state. The difference between these values seems to define an entanglement gap for the Jain state itself, although larger sizes are necessary to verify this.

The presence of an entanglement gap in the Jain state differentiates it from ``pure'' CFT states, and makes the counting of the edge-state spectrum difficult. To proceed, we count only the eigenvalues of the Jain state that match the eigenvalues of the Coulomb spectrum (below the green line in Fig[\ref{entcoulomb}]). This should provide us with the ``universal'' counting of edge states for our finite size-system. As seen in Fig[\ref{entcoulomb}], this counting is $1:2:5$ for $\Delta_L=0,1,2$. This matches the counting of two $U(1)$ free bosons which is also the prediction of the hierarchy or Composite Fermion construction.

The Coulomb state follows most of the low-energy eigenvalues of the Gaffnian (up to $\xi=8$) and Jain state (up to $\xi=10$).  While two states that have almost identical spectral decomposition necessarily have large overlap, the converse is not true, as large overlap can be accidental. Our finding of almost identical low-energy spectral decompositions indicates that the large overlap these three states show with each other is not accidental, which is very puzzling considering the Gaffnian and Jain states represent different states of matter. In order to appreciate the similar structure of these states, we plot the overlap of several reduced density matrix eigenstates of the Gaffnian with Coulomb  for different $L_z^{(A)}$, and find Fig.[\ref{entcoulomb}], nearly perfect overlap up to the point where the system is driven into a phase transition by the addition of negative delta-function potential.

\begin{figure}
\includegraphics[width=3.5in]{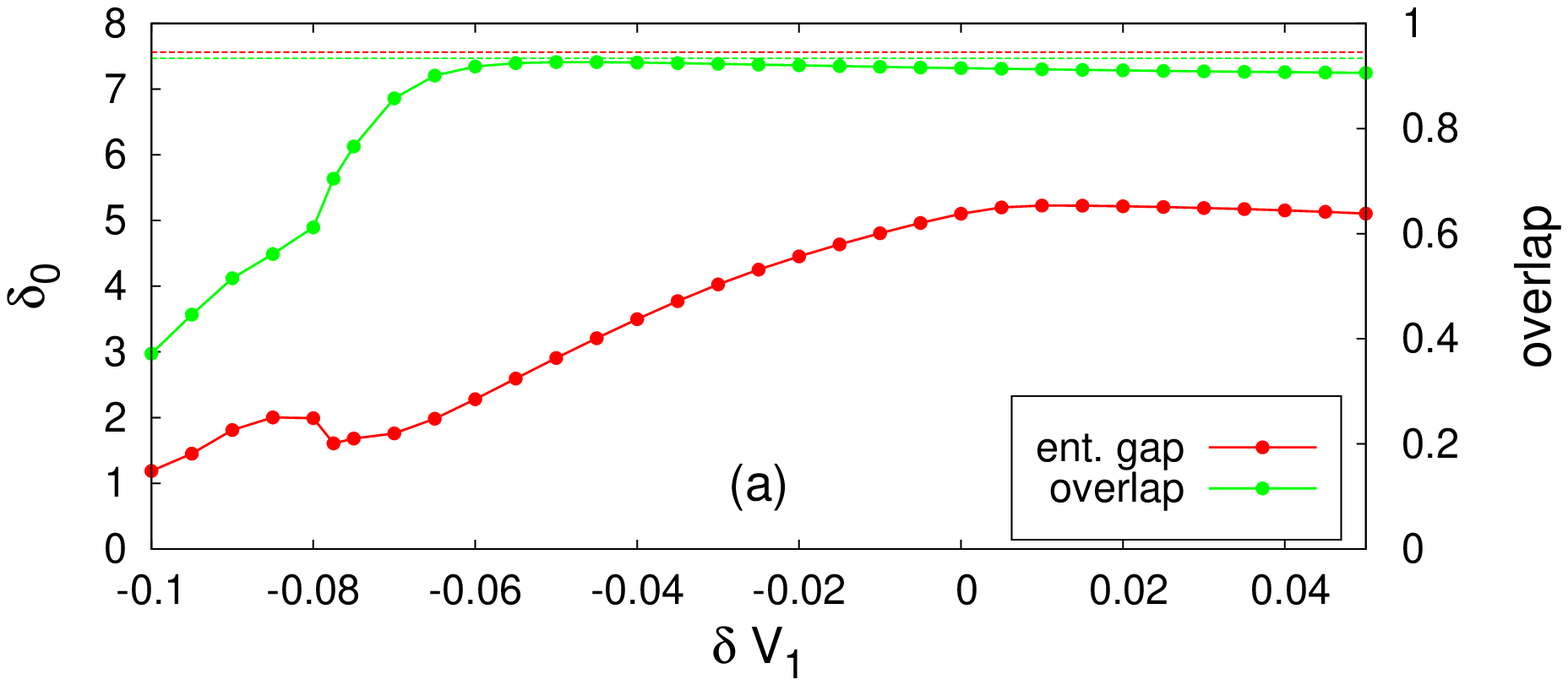}
\includegraphics[width=3.5in]{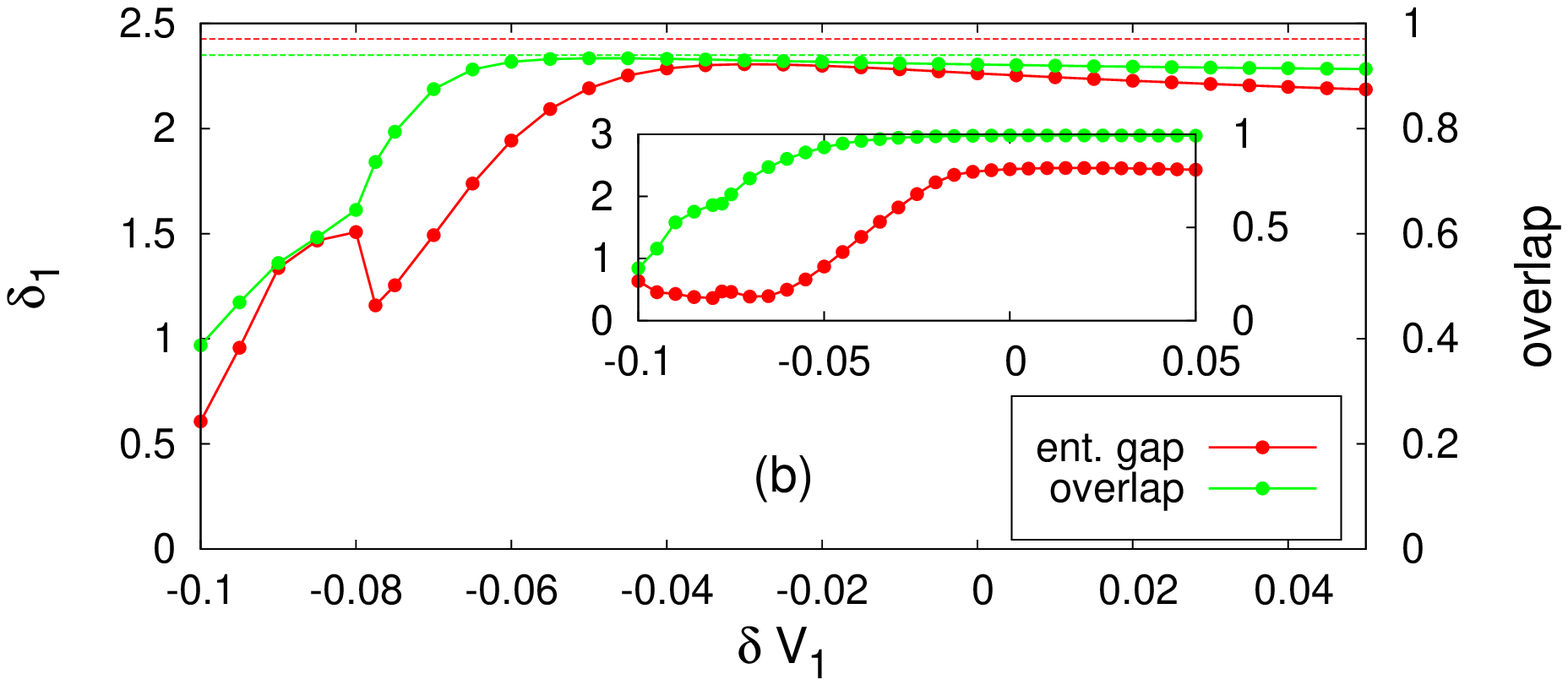}
\includegraphics[width=3.5in]{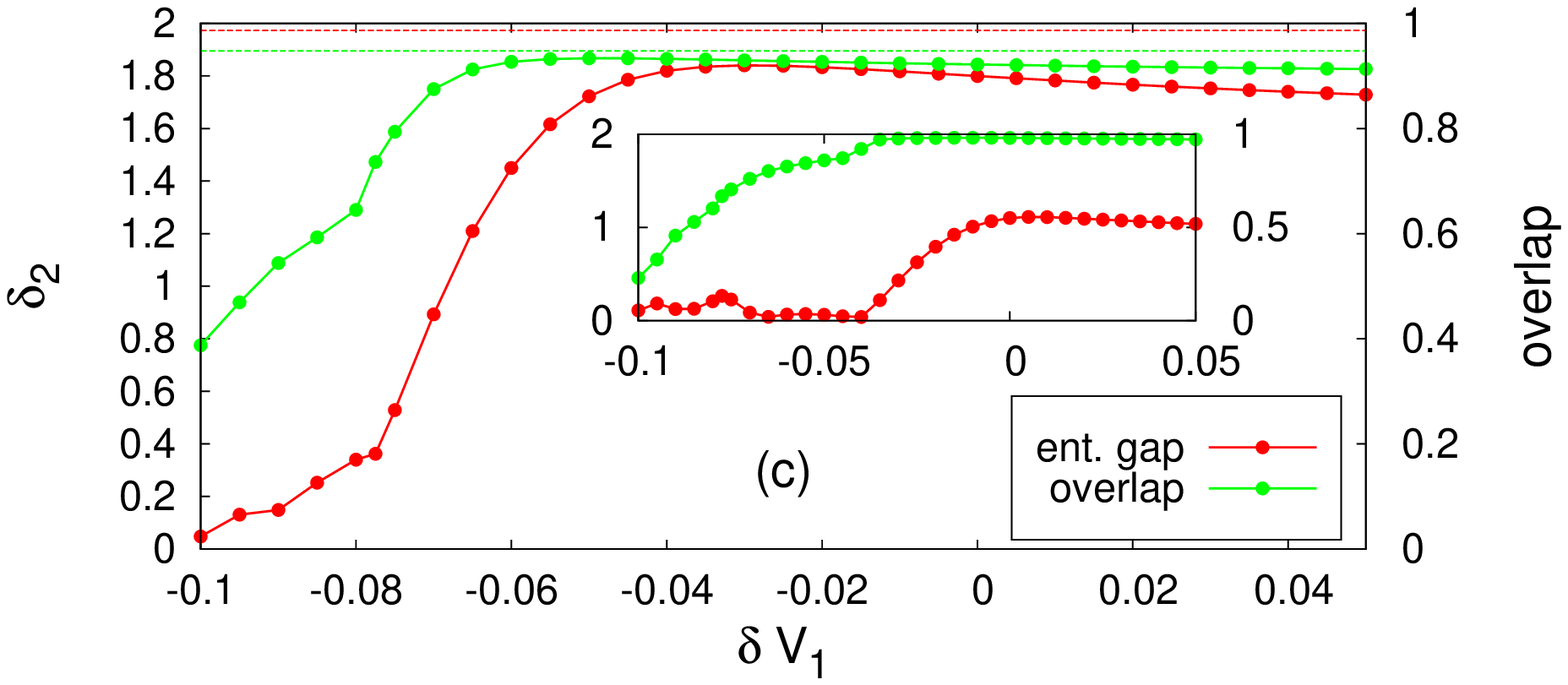}
\caption{ {\it Red plots} : Entanglement gap of the coulomb state for 3 different values of $L_z^{(A)} =80$ (Fig (a) top), $L_z^{(A)} =79$ (Fig (b) middle), $L_z^{(A)} =78$ (Fig (c) bottom) as a function of added hard-core potential. The entanglement gap is discontinuous (a and b) or almost vanishing (c) at roughly the same values of $\delta V_1$ for which the overlap of the Gaffnian and Jain with the Coulomb ground-state also collapses (see inset Fig.[\ref{overlaps}]). The red dotted line is the entanglement gap for the Jain state. {\it Green plots}: Overlap of the reduced density matrix eigenstates for each of the $L_z^{(A)} =80,79,78$ between Coulomb and Gaffnian. The green dotted line is the calculation result between Jain and Gaffnian.  {\it Insets} : Similar results in the two U(1) free boson sector.}\label{entgap}
\end{figure}

As in \cite{li2008}, we denote the gap between the lowest
two $\xi_i$, at the $L_z^{(A)}$ value where the highest-$L^{(A)}_z$ member
of the CFT spectrum occurs, as $\delta_0$. In Fig.[\ref{entgap}], this is the
gap between the lowest two states at $L^{(A)}_z = 80$. We define the quantities $\delta_{1,2}$ \cite{li2008}, as the gaps at
$L^{(A)}_z=79,78$ values between the values of the $\xi_i$'s for the CFT state and the next Coulomb value. As noted previously, the Jain state has  its own entanglement gap, equal at $L_z^{(A)}$  with the difference between the lowest $\xi_i \approx 6$ and the next one at $\xi_i \approx 13.5$ We study what happens to the spectrum, in particular to the entanglement gaps $\delta_{0,1,2}$ as
we tune the interaction away from the FQH state across
a quantum phase transition. In figure 6 (lower panel), we plot $\delta_0$ as a function of the
pseudopotential $\delta V_1$ for the $\nu = 2/5$  case. This clearly
shows a dramatic decrease of the ``entanglement gap''
around the region of the phase transition. For values of $\delta V_1 < 0.08$ the CFT-like structure
of the entanglement spectrum is lost. We also investigate the dependence of the entanglement gap with system size and conjecture that it remains finite in the thermodynamic limit.  Entanglement gaps can also be computed in the two U(1) free boson sector (see insets of  Fig.[\ref{entgap}]). Both $\delta_1$ and $\delta_2$ gaps close or become negligible for values of $\delta V_1$ which are larger than those involved in the gaffnian sector. We notice that for $N=12$, this seems to be correlated to the first excited state having its angular momentum changing from $L=6$ to $L=2$ (close to $\delta V_1=-0.06$).

In conclusion, we analyzed the topological structure of Jain states focusing on the $\nu=2/5$ state and compared it to the Coulomb ground-state and to a recent non-unitary state at the same filling factor and shift. We showed that the Jain states at filling $k/(2k+1)$ exhibit a squeezing property that severely reduces the size of the Hilbert space needed to construct them. We found the structure of zeroes of these states and showed that they are zero modes (but not highest density) of a $k+1$-body pseudopotential. We showed that fermionic Jain states are a direct product of a Vandermonde determinant and a symmetric polynomial vanishing as the second power of the difference in coordinates of a $k+1$-particle cluster. We analyzed the entanglement spectrum of Jain, Coulomb and non-unitary Gaffnian state at $\nu=2/5$ and found a similar low-energy structure which proves that their large overlap is not accidental; the entanglement gap remains finite in the thermodynamic limit, but the Jain state contains some ``low-energy'' levels which are not included in the Gaffnian.  Nevertheless, the similarity of their entanglement spectra is puzzling as the Jain and Gaffnian (Jack) state correspond, in the thermodynamic limit, to different states of matter, one gapped and the other conjectured to be gapless \cite{read2008}. The overlap of both states with the Coulomb groundstate vanishes upon the addition of a hard-core attraction which drives the Coulomb system through a phase transition.

 B.A.B. also wishes to thank E. Rezayi, S. Simon, P. Bonderson for discussions. N.R. is grateful to E. Bergholtz for useful comments.

\end{document}